\documentclass[reprint, superscriptaddress, secnumarabic, amssymb, nobibnotes, aps, prl]{revtex4-1}

\usepackage{graphicx}
\usepackage{epstopdf}
\usepackage[T1]{fontenc}
\usepackage[latin9]{inputenc}
\usepackage{amsbsy}
\usepackage{gensymb}
\usepackage[T1]{fontenc}
\usepackage[latin9]{inputenc}
\usepackage{amsmath}
\usepackage{amssymb}
\usepackage{bbm}
\usepackage{braket}
\usepackage{xcolor}
\allowdisplaybreaks
\usepackage{graphicx}
\usepackage[colorlinks=true]{hyperref}  
\hypersetup{
    bookmarks=true,         
    unicode=false,          
    pdftoolbar=true,        
    pdfmenubar=true,        
    pdffitwindow=false,     
    pdfstartview={FitH},    
    pdftitle={Enhancement of the superconducting transition temperature by Re doping},    
    pdfauthor={D. Singh, R. P. Singh},     
    pdfsubject={},   
    pdfcreator={},   
    pdfproducer={}, 
    pdfkeywords={} {} {}, 
    pdfnewwindow=true,      
    colorlinks=true,       
    linkcolor=blue, 
    citecolor=blue,        
    filecolor=magenta,      
    urlcolor=blue           
} 
\usepackage[normalem]{ulem}

\newcommand{\figref}[1]{Fig.~\ref{#1}}

\begin{document}
\title{\textrm{Enhancement of the superconducting transition temperature by Re doping in Weyl semimetal MoTe$_{2}$}}
\author{Manasi Mandal}
\affiliation{Indian Institute of Science Education and Research Bhopal, Bhopal, 462066, India}
\author{Sourav Marik}
\affiliation{Indian Institute of Science Education and Research Bhopal, Bhopal, 462066, India}
\author{K. P. Sajilesh}
\affiliation{Indian Institute of Science Education and Research Bhopal, Bhopal, 462066, India}
\author{Arushi}
\affiliation{Indian Institute of Science Education and Research Bhopal, Bhopal, 462066, India}
\author{Deepak Singh}
\affiliation{Indian Institute of Science Education and Research Bhopal, Bhopal, 462066, India}\
\author{Jayita Chakraborty}
\affiliation{Indian Institute of Science Education and Research Bhopal, Bhopal, 462066, India}
\author{Nirmal Ganguli}
\affiliation{Indian Institute of Science Education and Research Bhopal, Bhopal, 462066, India}
\author{R. P. Singh}
\email[]{rpsingh@iiserb.ac.in}
\affiliation{Indian Institute of Science Education and Research Bhopal, Bhopal, 462066, India}

\date{\today}
\begin{abstract}
\begin{flushleft}

\end{flushleft}
This work presents the emergence of superconductivity in Re substituted topological Weyl semimetal MoTe$_{2}$. Re substitution for Mo sites lead to a sizable enhancement in the superconducting transition temperature (T$_{c}$). A record high T$_{c}$ at ambient pressure in a 1T$'$-MoTe$_{2}$ (room temperature structure) related sample is observed for the Mo$_{0.7}$Re$_{0.3}$Te$_{2}$ composition (T$_{c}$ = 4.1 K, in comparison MoTe$_{2}$, shows a T$_{c}$ of 0.1 K). The experimental and theoretical studies indicate that Re substitution is doping electrons and facilitates the emergence of superconductivity by enhancing the electron-phonon coupling and density of states at the Fermi level. Our findings, therefore, open a new way to further manipulate and enhance the superconducting state together with the topological states in 2D van der Waals materials.  
\end{abstract}
\maketitle
\section{Introduction}

The exotic quantum states such as topological states \cite{1,2,3}, Dirac semimetals \cite{4,5,6,7} and Weyl semimetals \cite{8,9,10} have been extensively studied in recent years as the new frontier of solid-state physics and chemistry. In this context, two-dimensional (2D) transition metal dichalcogenides (TMDs, TMCh$_{2}$) have attracted tremendous interest due to their rich structural and intercalation chemistry, unusual quantum states, and promising potential applications \cite{11,12,13,14,15}. Illustrative examples include topological field-effect transitions based on quantum spin Hall (QSH) insulators \cite{16}, extremely large magnetoresistance (in T$_{d}$ - WTe$_{2}$) \cite{13} and artificial van der Waals heterostructures \cite{17} with high on/off current ratio in TMDs. At the same time, rich electron coupling interactions foster exotic states of matter such as charge density waves (CDW) and unconventional superconductivity (SC) in TMDs \cite{18,19,20}.

\begin{figure}
\includegraphics[width=1.0\columnwidth]{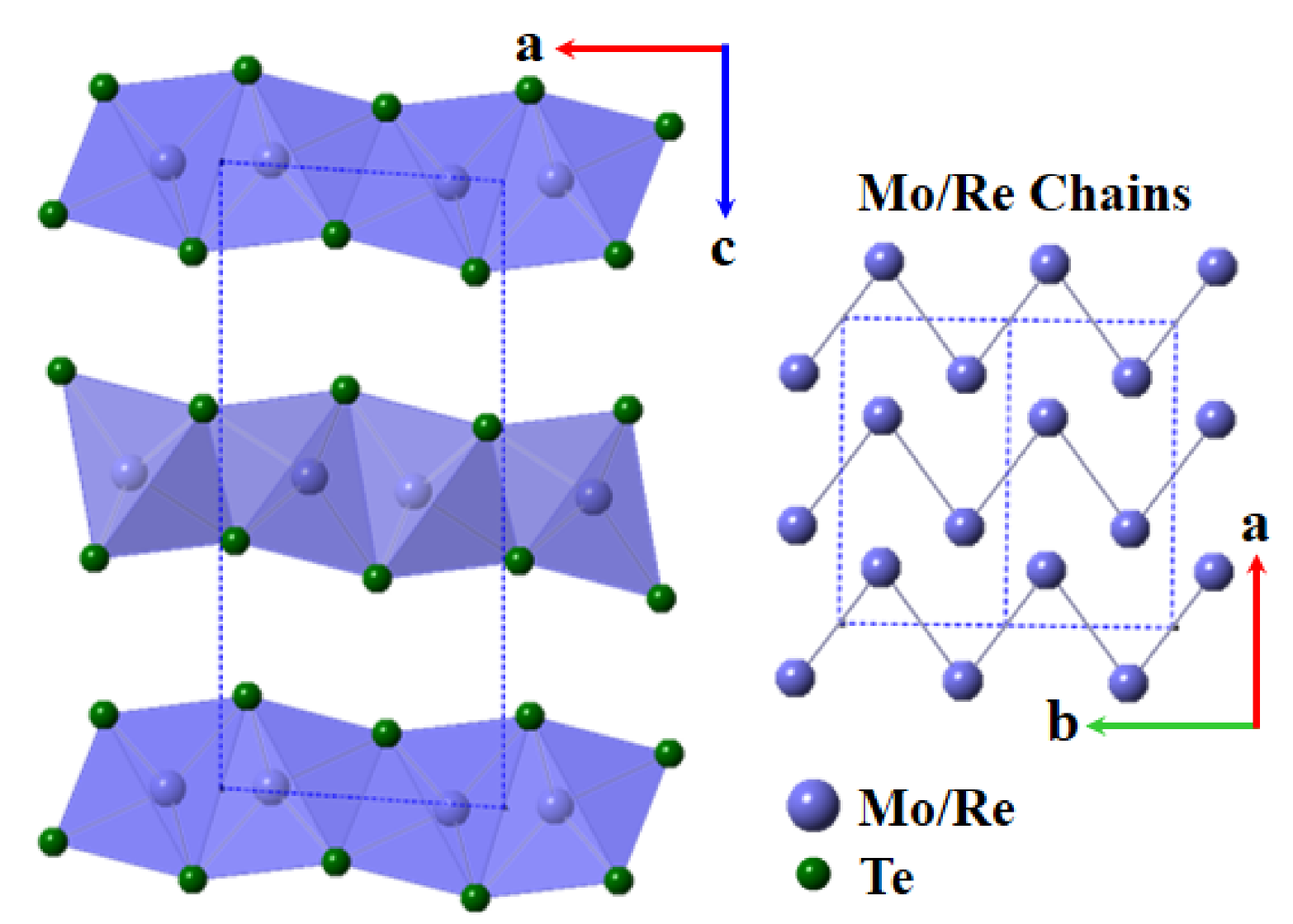}
\caption{\label{Fig1:STRUC} Room temperature crystal Structure of the Mo$_{1-x}$Re$_{x}$Te$_{2}$ materials (space group P21/m, 1T' phase). The zigzag Mo/Re chains along b direction are highlighted here.}
\end{figure}

The intrinsic chemical flexibility and the electron interactions achieved through the van der Walls stacking are enhanced further by the structural polymorphism of TMDs. MoTe$_{2}$ is a unique example among TMDs with the richest collections of variations on a structural theme (the semiconducting trigonal prismatically 2H phase, the semimetallic monoclinic 1T$'$ (\figref{Fig1:STRUC}) and orthorhombic T$_{d}$ phase) \cite{21,22,23}. It shows an inversion symmetric monoclinic (space group P21/m, 1T$'$ phase) - non - centrosymmetric orthorhombic (space group Pmn21, T$_{d}$ Phase) structural phase transition at 250 K. The orthorhombic T$_{d}$ phase with broken inversion symmetry is predicted to be a type - II Weyl semimetal and allows to study the fundamentals of novel quantum properties such as topological surface state Fermi arcs and chiral anomaly induced negative magnetoresistance \cite{24,25,26,27,28}. Very recently, superconductivity with a very low transition temperature (T$_{c}$) of   0.1 K was discovered in the orthorhombic T$_{d}$ MoTe$_{2}$ \cite{11}. However, interestingly the application of external physical pressure enhances the T$_{c}$ dramatically and further a dome - shaped superconducting phase diagram was established. Therefore, besides the exotic type - II Weyl semimetallic state, the discovery of superconductivity in T$_{d}$ MoTe$_{2}$ offers a fascinating opportunity to explore superconductivity and topological states by tuning the local structural distortion or by manipulating the chemical pressure.

On the lure of increasing the superconducting transition temperature and to find out the origin of superconductivity in MoTe$_{2}$, most of the studies were concentrated on the manipulation of chemical pressure by substituting the chalcogenide site (isovalent substitution). Indeed, significant enhancement of T$_{c}$ is achieved by replacing Te with S and Se (highest T$_{c}$ = 2.5 K for MoTe$_{2-x}$Se$_{x}$ series \cite{29} and 1.3 K for MoTe$_{2-x}$S$_{x}$ series \cite{30}), however, the choices are limited in this case. On the other hand, substitution on the Mo sites leads to the advent of several novel properties, such as an enhancement of thermopower near the critical region between the polar and nonpolar metallic phases was discovered in Mo$_{1-x}$Nb$_{x}$Te$_{2}$ (hole doping) \cite{31}. Furthermore, Mo$_{1-x}$W$_{x}$Te$_{2}$ alloy system is a promising candidate for the development of phase change memory technology \cite{32}. Nevertheless, superconductivity in Mo substituted MoTe$_{2}$ is still elusive.

In this paper, we show the emergence of superconductivity in Mo$_{1-x}$Re$_{x}$Te$_{2}$ phases. The dependence of T$_{c}$ with composition (x) and electron doping in the structure is described here. Furthermore, combining the ac and dc susceptibility, resistivity and specific heat measurements we demonstrate that Mo$_{0.7}$Re$_{0.3}$Te$_{2}$ shows the highest T$_{c}$ = 4.1 K at ambient pressure in a MoTe$_{2}$ related system.
\section{Experimental Details}

Pure phase polycrystalline samples of composition Mo$_{1-x}$Re$_{x}$Te$_{2}$ (x = 0, 0.05, 0.1, 0.2, 0.3 and 0.4) were prepared by the standard solid-state reaction process. Stoichiometric mixtures of Mo (99.9$\%$ pure), Re (99.99$\%$ pure) and Te (99.99$\%$ pure) powders were ground together and sealed in an evacuated quartz tube. The ampoules were first heated at 1100$°$C for 24 hours. The obtained samples were then ground, pelletized, and annealed at 1100$°$C for 24 hours, followed by ice water quenching to avoid the formation of the 2H phases. Single crystals of composition Mo$_{0.9}$Re$_{0.1}$Te$_{2}$ was grown via chemical vapour transport method using polycrystalline Mo$_{0.9}$Re$_{0.1}$Te$_{2}$ powder and iodine as a transport agent. Crystals were obtained by sealing 3g of Mo$_{0.9}$Re$_{0.1}$Te$_{2}$ powder and I$_{2}$ in a quartz tube. The tube was then flushed with Ar, evacuated, sealed and heated in a two-zone furnace. Crystallization was carried out from 1100$°$C (hot zone) to 1000$°$C (cold zone) for 14 days. Finally, the sealed quartz tube was quenched in ice water to avoid the formation of the hexagonal phase. Powder X-ray diffraction (XRD) was carried out at room temperature (RT) on a PANalytical diffractometer equipped with Cu-K$_{\alpha}$ radiation ($\lambda$ = 1.54056 \text{\AA}). Sample compositions were checked by a scanning electron microscope (SEM) equipped with an energy-dispersive X-ray (EDX) spectrometer. DC magnetization and ac susceptibility measurements were performed using a Quantum Design superconducting quantum interference device (MPMS 3, Quantum Design). The Hall measurements were carried out on the physical property measurement system (PPMS, Quantum Design, Inc.). Electrical resistivity measurements were performed on the PPMS by using a conventional four-probe ac technique at frequency 17 Hz and excitation current 10 mA. The measurements were carried out under the presence of different magnetic fields. Specific heat measurement was performed by the two tau time-relaxation method using the PPMS in zero magnetic field.

\section{Results and Discussion}

\begin{figure}
\includegraphics[width=1.0\columnwidth]{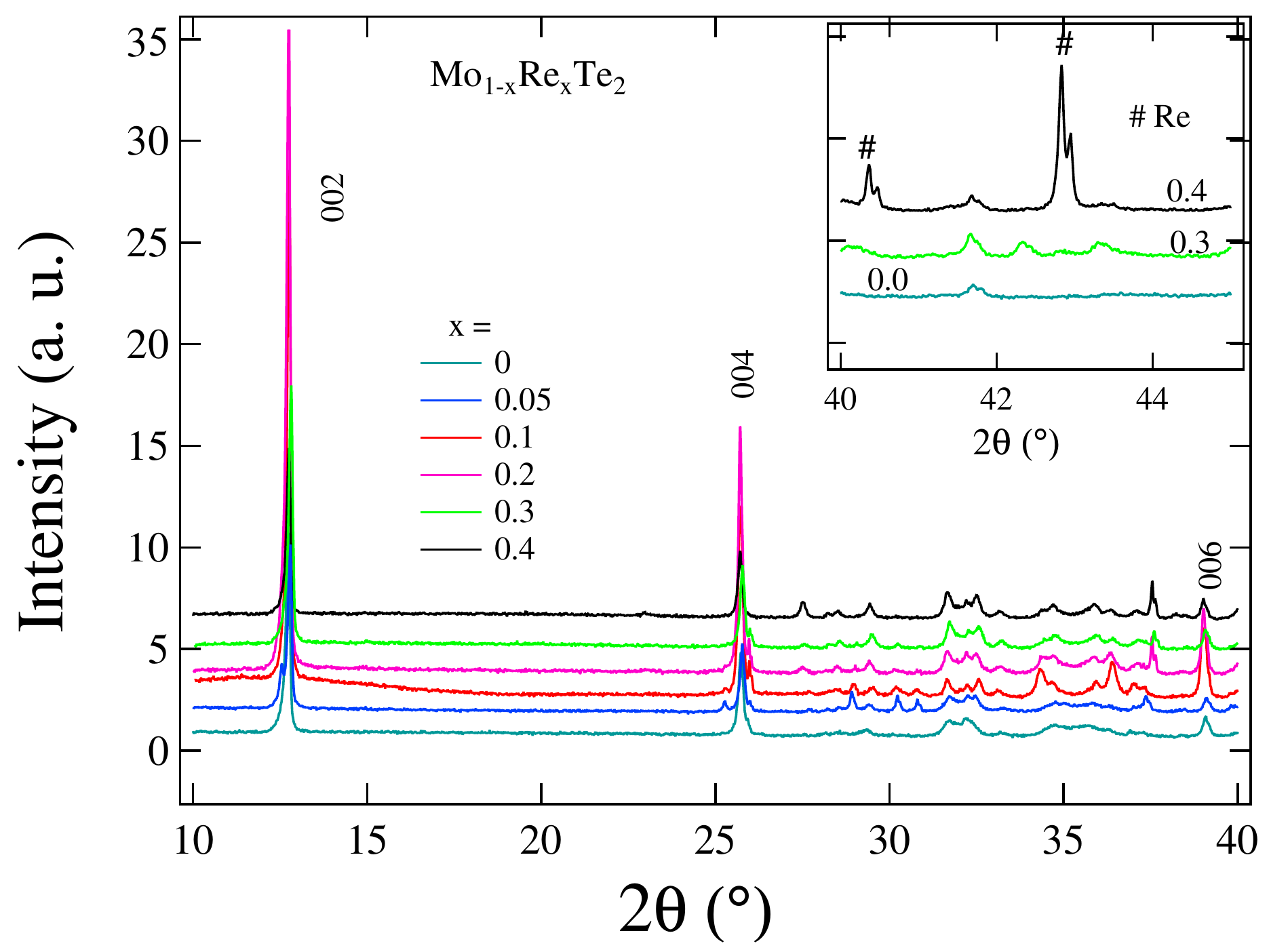}
\caption{\label{Fig2:XRD} Room temperature (RT) powder X-ray diffraction pattern for Mo$_{1-x}$Re$_{x}$Te$_{2}$ samples. Inset shows the unreacted Re peak in the XRD pattern for x = 0.4 sample.}
\end{figure}

\begin{figure}
\includegraphics[width=1.0\columnwidth]{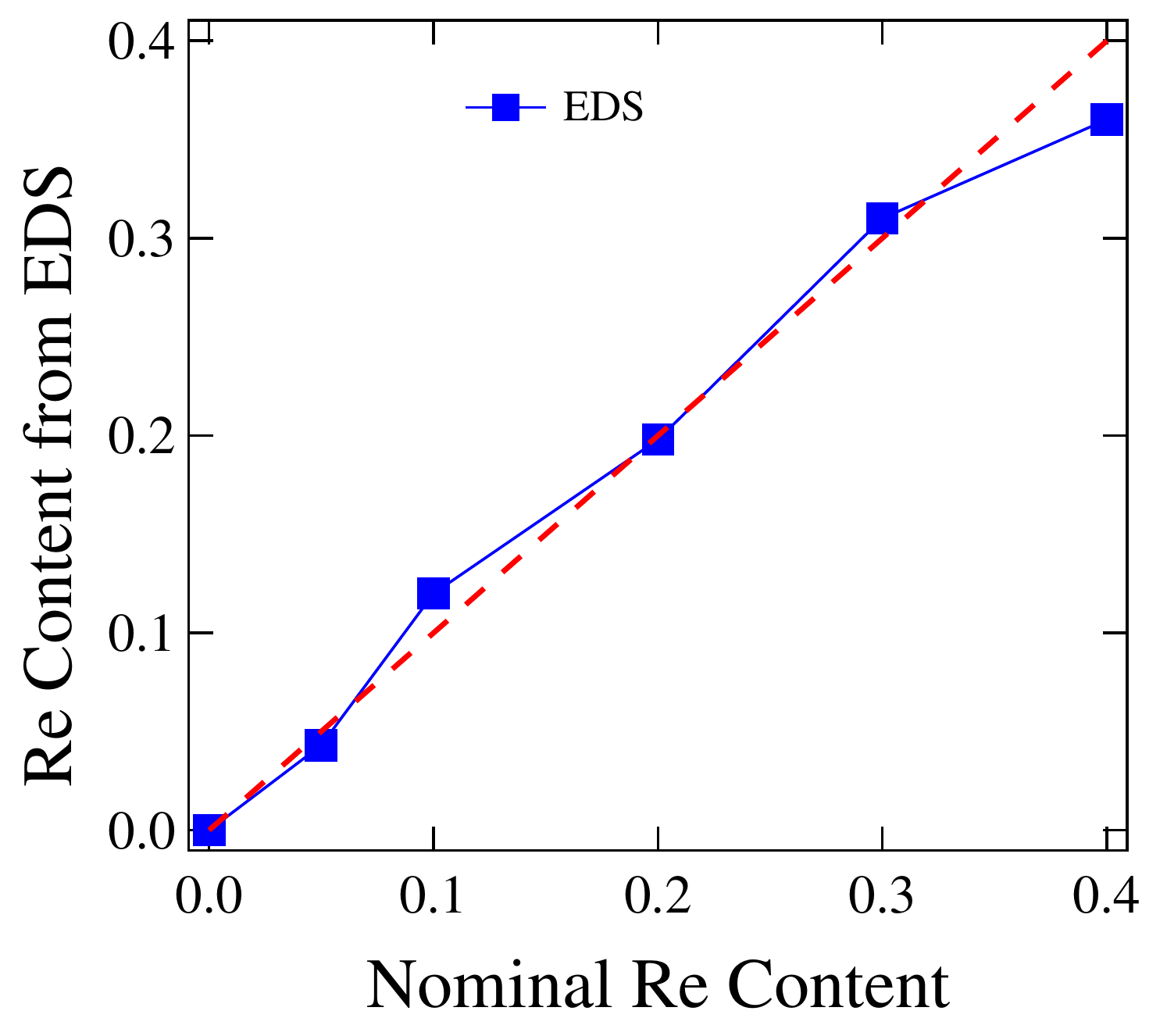}
\caption{\label{Fig3:EDS} Average Re concentration (average of 10 points) obtained from the EDX measurements for Mo$_{1-x}$Re$_{x}$Te$_{2}$ samples.}
\end{figure}

Room temperature (RT) powder X-ray diffraction (XRD) patterns (\figref{Fig2:XRD}) for the Mo$_{1-x}$Re$_{x}$Te$_{2}$ samples indicate that all the compounds can be isolated as single phase. However, an unreacted peak of Re in the RT XRD pattern is observed for the sample with x = 0.4 (inset in \figref{Fig2:XRD}), and the intensity of the peak increases with increasing Re content in the nominal composition. At room temperature, all the samples crystallize in a centrosymmetric monoclinic structure (CdI$_{2}$ type, P21/m space group) consisting of edge-sharing (Mo/Re)Te$_{6}$ octahedra with a strong distortion (\figref{Fig1:STRUC}). The Energy-dispersive X-ray spectroscopy (EDX) analysis of the Mo$_{1-x}$Re$_{x}$Te$_{2}$ samples confirm the existence of Re, Mo and Te in the samples. \figref{Fig3:EDS} highlights the obtained amount of the Re in the structure for all the samples. EDX measurements were carried out at ten different points for each sample. The average of 10 points for each sample is shown in the \figref{Fig3:EDS} and this follows the nominal Re content. Therefore, within our experimental limit, we can say that the Re distribution in the samples is quite homogeneous. However, for the x = 0.4 sample the actual Re concentration is found to be smaller than the nominal one. This, in fact, supports the observation of the unreacted peaks of Re in the RT XRD pattern for the x = 0.4 sample. 

\begin{figure}
\includegraphics[width=1.0\columnwidth]{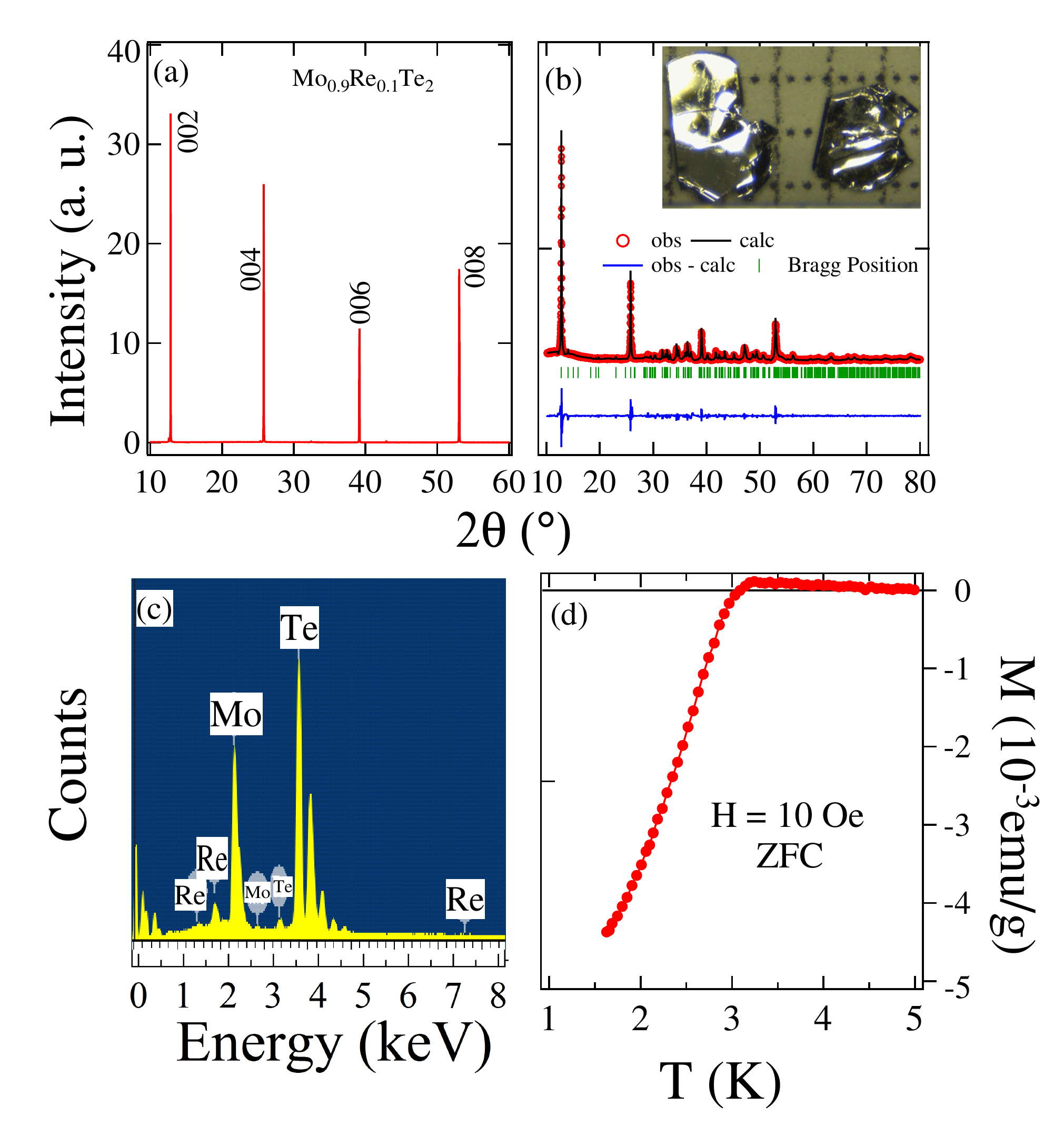}
\caption{\label{Fig4:XRDCRYS} (a) RT XRD pattern on the cleaved single crystal (b) the Rietveld refinement plots (c) EDX pattern and (d) temperature variation of the magnetic moment for Mo$_{0.9}$Re$_{0.1}$Te$_{2}$ crystals. Images of the grown crystals are shown in the inset of Fig 4(b).}
\end{figure}

Also, we have grown the single crystal for Mo$_{0.9}$Re$_{0.1}$Te$_{2}$ composition. \figref{Fig4:XRDCRYS} (a) shows the RT XRD pattern on the naturally cleaved surface of the single crystal ([00l] reflections). \figref{Fig4:XRDCRYS} (b) shows the final plot of the Rietveld refinement of the powder XRD for Mo$_{0.9}$Re$_{0.1}$Te$_{2}$ crystals. EDX result shows that the average composition of our grown crystal is Mo$_{0.91}$Re$_{0.09}$Te$_{2}$. Temperature variation of the magnetic measurements on the crystal highlights superconducting transition at 3.2 K. Nevertheless, due to the low superconducting volume fraction in the crystals and the difficulty in processing (due to the small size), we have used the polycrystalline data in rest of the paper.     

\begin{figure}
\includegraphics[width=0.85\columnwidth]{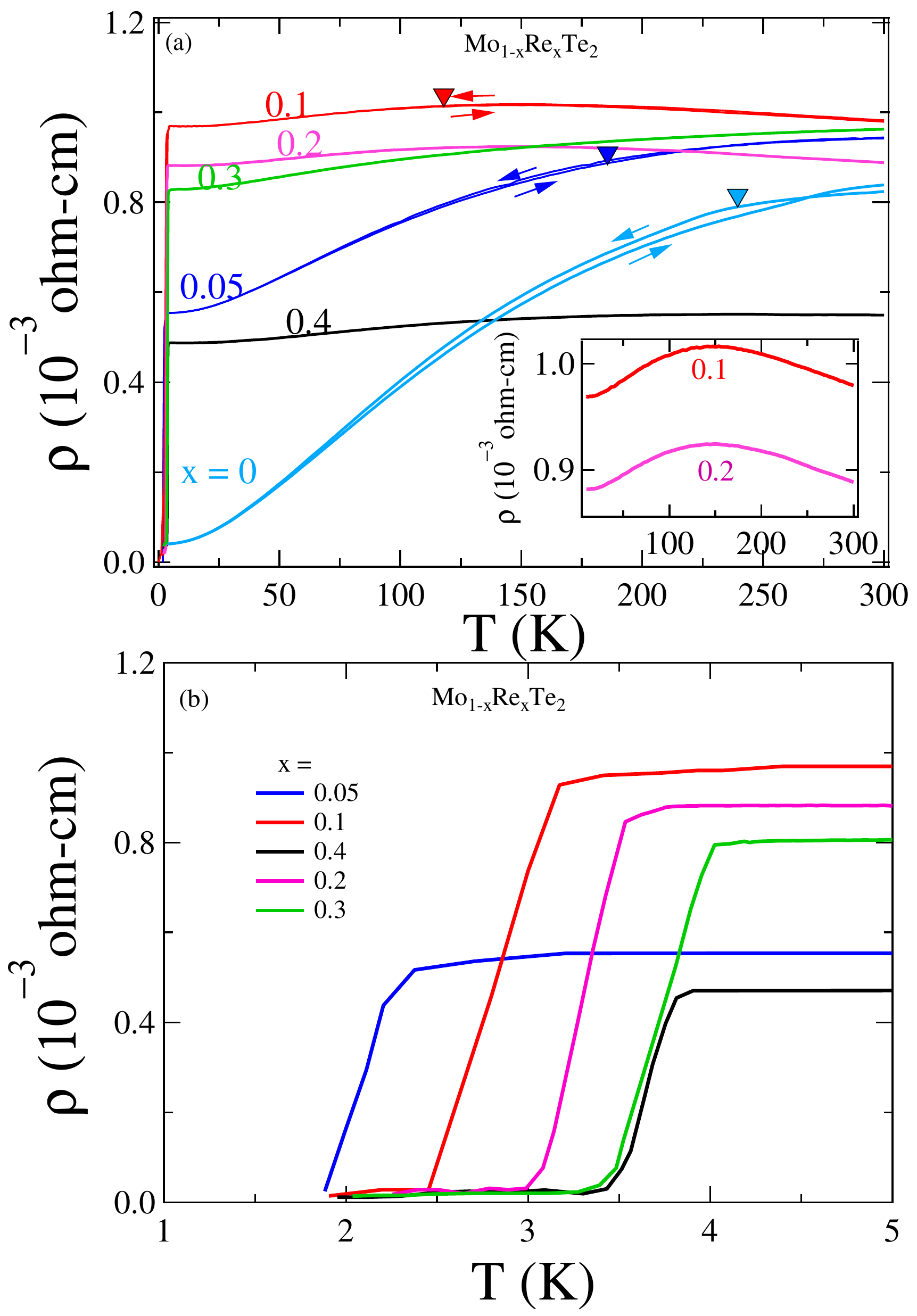}
\caption{\label{Fig5:RESMAG} Temperature dependence of (a) resistivity under zero magnetic field for Mo$_{1-x}$Re$_{x}$Te$_{2}$ samples. Solid triangles indicate the structural transition temperatures. Arrows highlight the cooling and heating process. Inset shows the observed anomaly in the normal state resistivity data for x = 0.1 and 0.2 samples. (b) Enlarged part of the resistivity measurements highlighting the superconducting transitions for Mo$_{1-x}$Re$_{x}$Te$_{2}$.}
\end{figure}

\begin{figure}
\includegraphics[width=1.0\columnwidth]{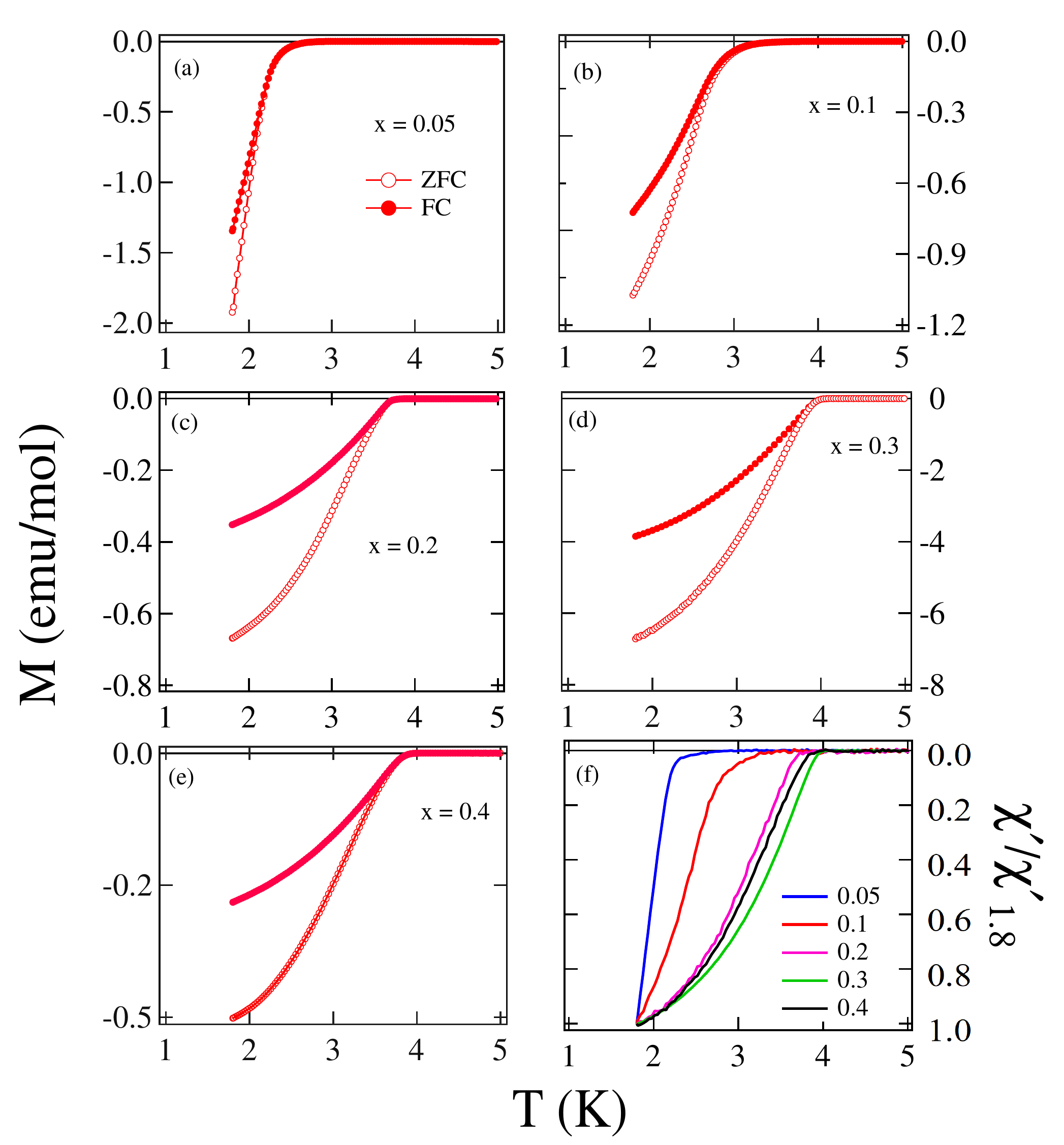}
\caption{\label{Fig6:MAG} Temperature dependence of (a)-(e) dc magnetization and (f) normalized ac susceptibility at a dc magnetic field of 10 Oe (ac excitation field = 2 Oe) for Mo$_{1-x}$Re$_{x}$Te$_{2}$ samples.}
\end{figure}

\begin{figure}
\includegraphics[width=1.0\columnwidth]{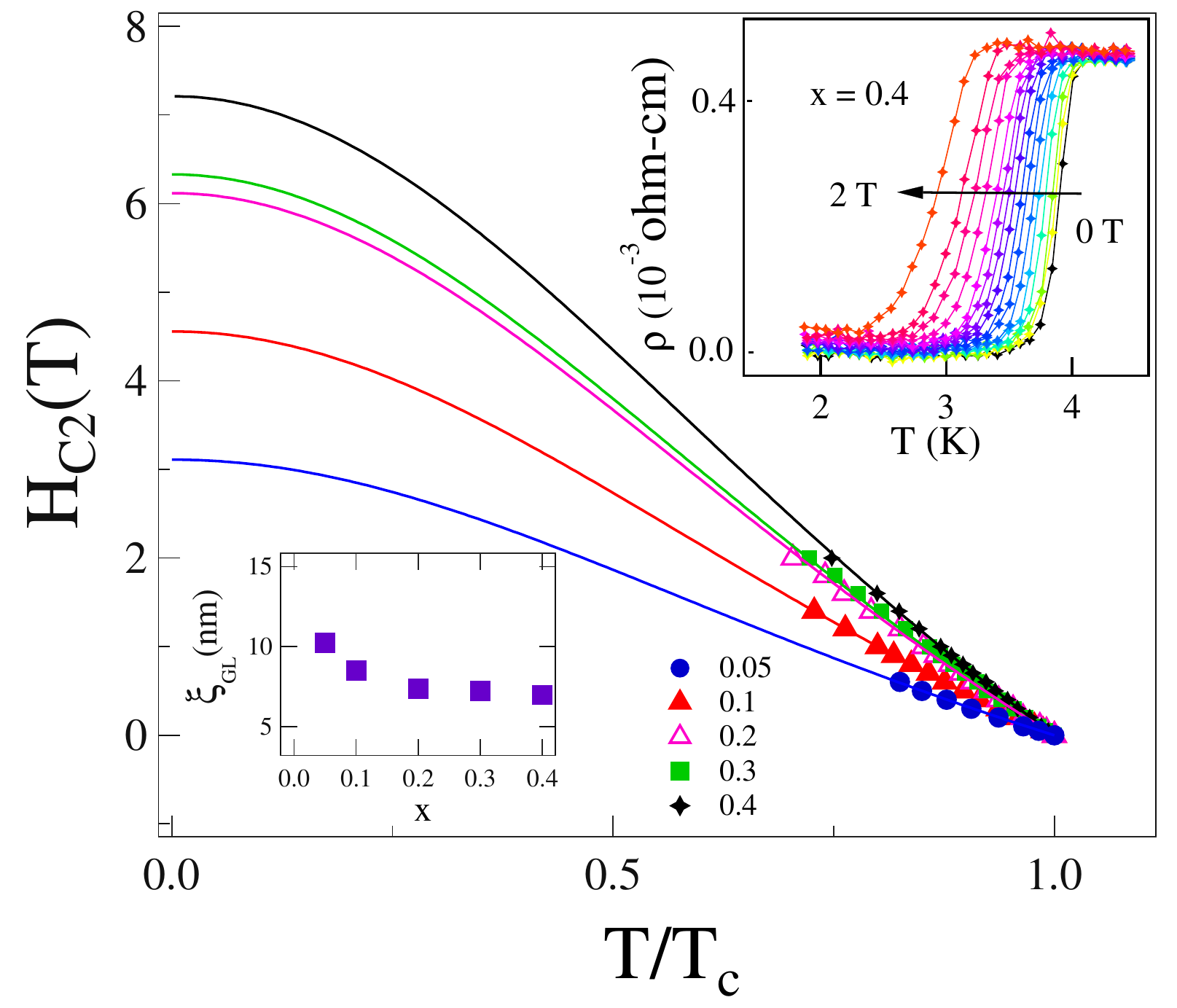}
\caption{\label{Fig7:HC2} Determination of the upper critical field (H$_{C2}$ (0)) using resistivity measurements. The solid lines represent the Ginzburg Landau fitting. The upper inset shows the magnetic field variation of the resistivity for x = 0.4 sample. The Ginzburg Landau coherence length for the Mo$_{1-x}$Re$_{x}$Te$_{2}$ materials are shown in the lower inset.}
\end{figure}

\figref{Fig5:RESMAG} shows the temperature dependence of resistivity for all the samples. The pristine MoTe$_{2}$ sample does not show superconducting state down to 1.8 K (T$_{d}$ MoTe$_{2}$ shows SC at 0.1 K \cite{11}), whereas the x = 0.05 sample exhibits a clear signature of superconducting state at 2.3 K. Interestingly, with increasing Re content in the structure, the superconducting transition temperature increases dramatically. Mo$_{0.7}$Re$_{0.3}$Te$_{2}$ (x = 0.3) shows a record high T$_{c}$ of 4.1 K in any MoTe$_{2}$ related sample at ambient pressure. However, beyond this compositional point, T$_{c}$ starts to decrease, and x = 0.4 sample shows lower onset T$_{c}$ than x = 0.3. At the same time, the normal state resistivity highlights an anomalous behavior (broad hump) for the samples with x = 0.2 and 0.3 (inset in \figref{Fig5:RESMAG} (a)). The presence of a hump type anomaly in the resistivity data is not surprising for TMDs and may indicates the presence of a CDW state \cite{33,34,35}. For instance, under high external pressure MoS$_{2}$ shows a similar hump type anomaly in the resistivity data due to the existence of the CDW state \cite{35}. Therefore, the observed anomaly in the normal state resistivity in our samples could be an indication of the appearance of the charge density wave (CDW) in the material. However, with further increasing the Re content in the structure metallic- type normal state behavior is observed. To verify the presence of CDW in the present materials, further experiments are required. The polar structural transition temperature (triangles in the \figref{Fig5:RESMAG} (a)), which is defined as the irreversibility (and an anomaly) between cooling and heating cycle of the resistivity data is found to decrease with increasing Re concentration. However, the irreversibility and the structural anomaly get weaker with Re doping and almost disappeared for samples with higher Re concentration (x$\geq$0.1). Therefore, based on the present resistivity data it is not possible to precisely determine the Re-doping evolution of the structural transition.      
The enhanced superconducting states are characterized by dc and ac (\figref{Fig6:MAG}) susceptibility measurements. Both the ac and dc susceptibility measurements confirm the evidence of bulk type - II superconductivity in Re doped MoTe$_{2}$. The variation and the values of the T$_{c}$ are consistent with the resistivity measurements. The lower critical field (H$_{C1}$) estimated from the isothermal magnetization curves are in the range of 20-50 Oe. Detailed study of the superconducting properties of the Mo$_{1-x}$Re$_{x}$Te$_{2}$ samples were carried out by measuring the resistivity and dc susceptibility under various applied magnetic fields (H). The T$_{c}$ is gradually suppressed with increasing magnetic field for all the samples (upper inset in \figref{Fig7:HC2}). The temperature variation of the upper critical field (H$_{C2}$) for the samples is shown in \figref{Fig7:HC2}. The critical temperatures are derived from the midpoint values of the superconducting transition in the resistivity measurements. The experimental H$_{C2}$ can be described by the Ginzburg Landau expression. 
\begin{equation}
H_{c2}(T) = H_{c2}(0)\frac{(1-(T/T_{c})^{2})}{(1+(T/T_{c})^2)}
\label{eqn1:hc2}
\end{equation}
The estimated values of the upper critical field (H$_{C2}$(0)) show a gradual increment with increasing Re content in the structure. Interestingly, the upper critical field (H$_{C2}$(0) = 7.2 T) for the Re rich composition Mo$_{0.6}$Re$_{0.4}$Te$_{2}$ is found to be very close to the BCS weak coupling Pauli limit (7.23 T). In comparison, for the pristine MoTe$_{2}$, H$_{C2}$(0) value is far below the Pauli limit \cite{11}. Therefore, the observed H$_{C2}$(0) value for Mo$_{0.6}$Re$_{0.4}$Te$_{2}$, which is almost equal to the Pauli limit may points to the unconventional nature of the super-conductivity in the sample. The variation of Ginzburg Landau coherence lengths $\xi_{GL}$(0), estimated from the following relation, as a function of Re concentration is shown as an inset (lower inset) in \figref{Fig7:HC2}. 
\begin{equation}
H_{c2}(0) = \frac{\Phi_{0}}{2\pi\xi_{GL}^{2}}
\label{eqn2:up}
\end{equation}
The calculated values of $\xi_{GL}$(0) are in the range of 6-11 nm and comparable with the $\xi_{GL}$(0) value obtained for the pristine MoTe$_{2}$ under 11.2 GPa \cite{11}.
\begin{figure}
\includegraphics[width=1.0\columnwidth]{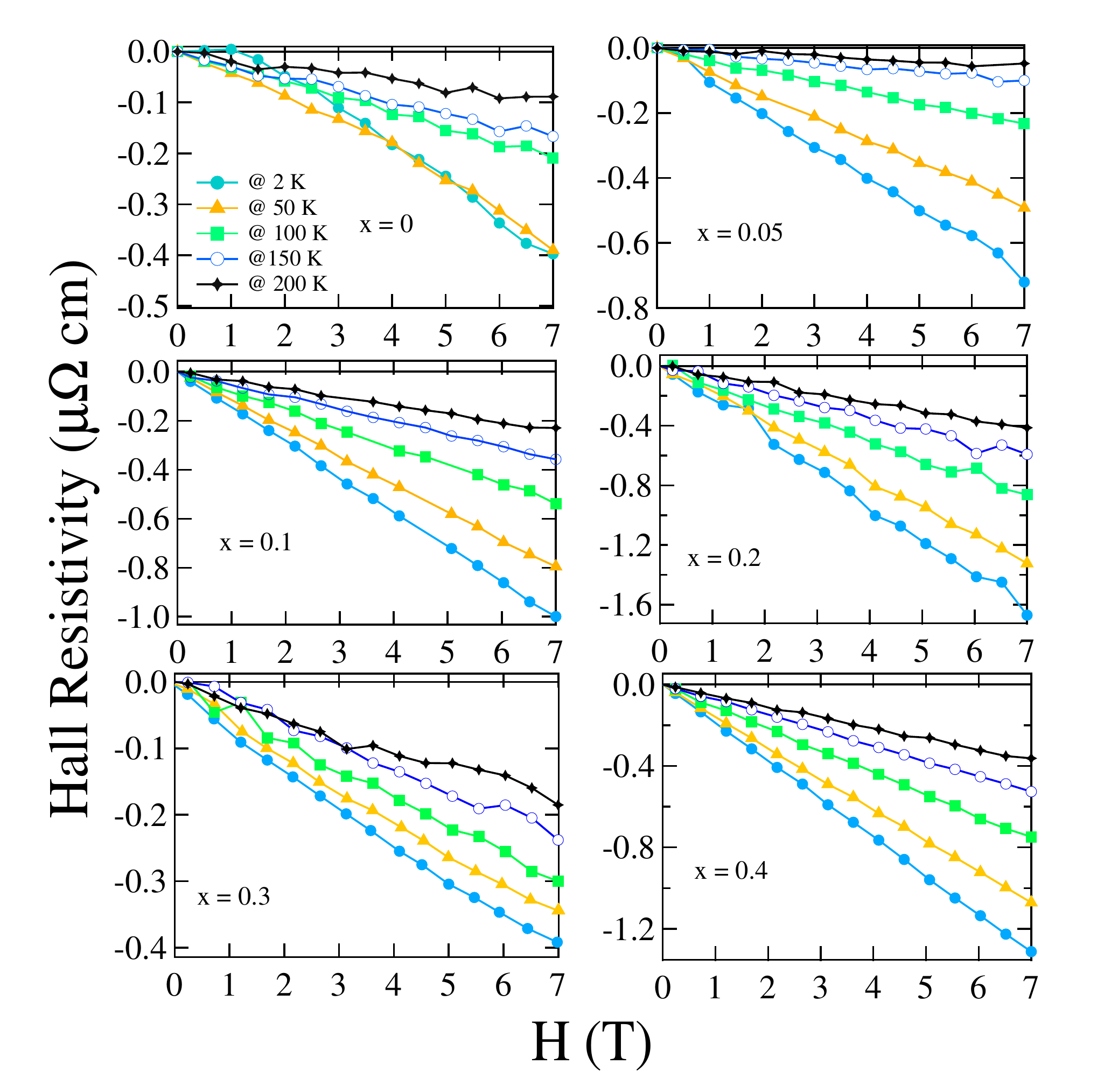}
\caption{\label{Fig8:HALL} Magnetic field variation of the Hall resistance at different temperatures for the Mo$_{1-x}$Re$_{x}$Te$_{2}$ samples.}
\end{figure}

\begin{figure}
\includegraphics[width=1.0\columnwidth]{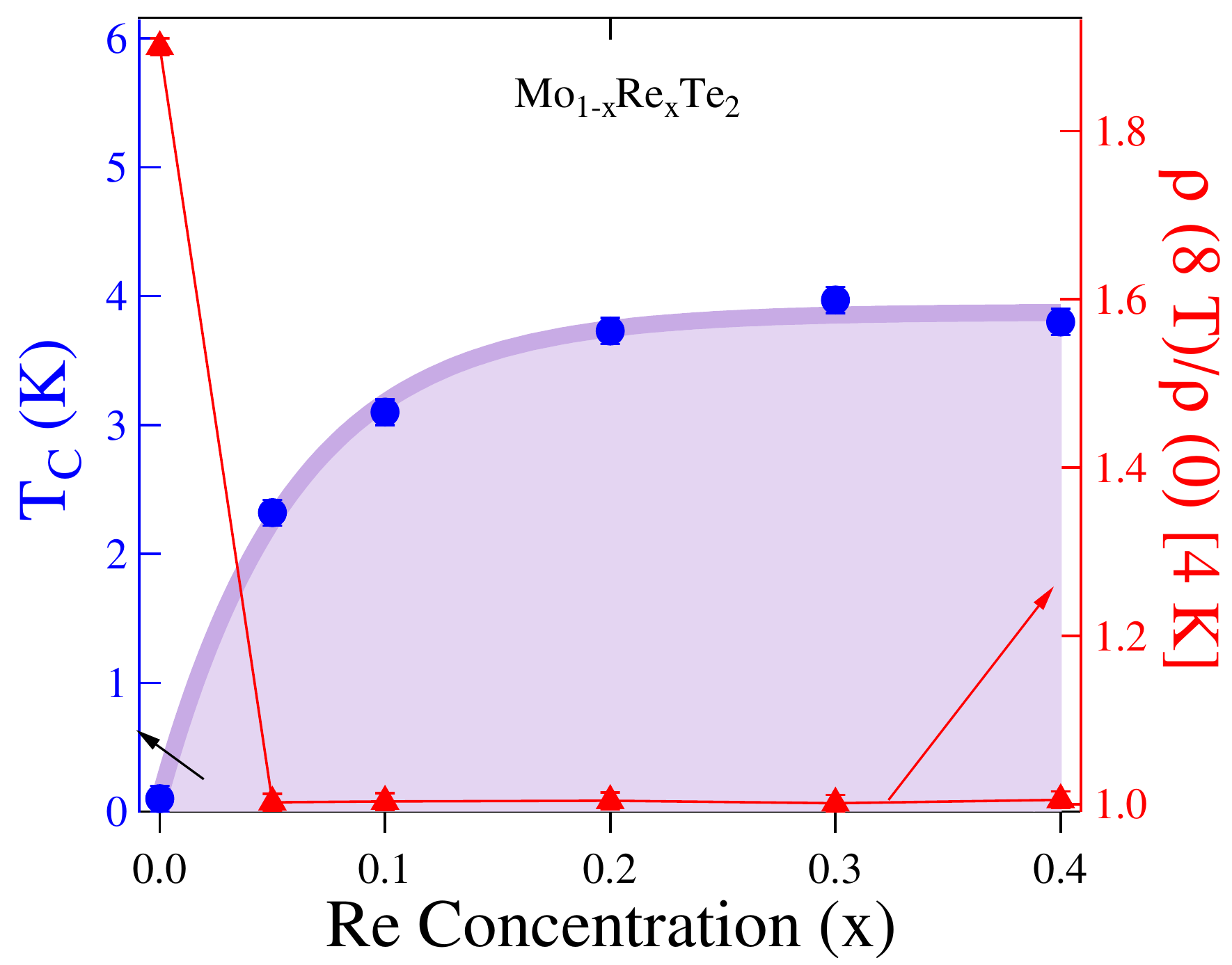}
\caption{\label{Fig9:ALL} The variation of the superconducting transition temperatures (T$_{c}$, left axis) and magnetoresistance (right axis) with Re substitution level in the structure for Mo$_{1-x}$Re$_{x}$Te$_{2}$.}
\end{figure}

To determine the type of the charge carriers and its evolution with the Re substitution, we have measured the Hall resistivity for all the samples (\figref{Fig8:HALL}). For the pristine MoTe$_{2}$ sample the magnetic field dependence of the Hall resistivity is curved with a negative slope, and this is typical of a semimetal having both hole and electron contributions. However, with increasing Re content in the material, the Hall profiles exhibit a straight-like behavior. This indicates that electron-like carriers are increasing with Re substitution. At the same time, a dramatic suppression of the magnetoresistance (MR, Fig. S3) is observed with Re substitution. It is noteworthy that in WTe$_{2}$, superconductivity emerges due to the suppression of large magnetoresistance by the application of external pressure \cite{36}. Upon increasing external pressure in WTe$_{2}$, the electron concentrations start increasing and at a critical pressure of 10.5 GPa, superconductivity emerges with a change in the sign of the Hall coefficient (positive to negative) \cite{36}. \figref{Fig9:ALL} summarizes the T$_{c}$ and MR values for Mo$_{1-x}$Re$_{x}$Te$_{2}$ as a function of Re concentration. 
This unambiguously highlights that the T$_{c}$ dramatically increases with Re substitution of the Mo sites in MoTe$_{2}$. On the contrary, Nb substituted MoTe$_{2}$ where Nb substitution induces holes in the structure do not exhibit superconductivity \cite{31}. 
\begin{figure}
\includegraphics[width=1.0\columnwidth]{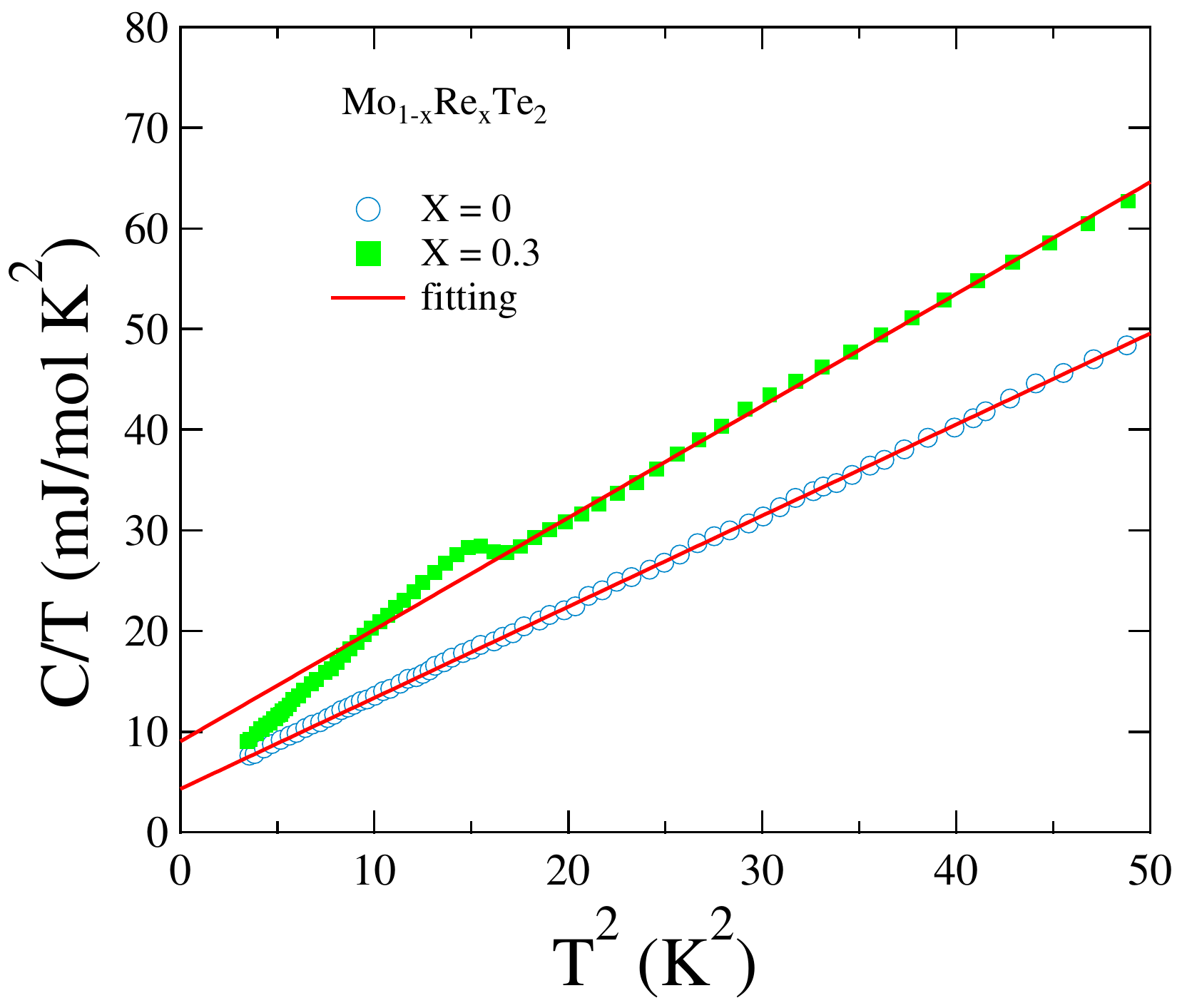}
\caption{\label{Fig10:CP} The low-temperature specific heat data for MoTe$_{2}$ and for Mo$_{0.7}$Re$_{0.3}$Te$_{2}$. The specific heat data above T$_{c}$ is fitted to the Debye model and shown by a solid line.} 
\end{figure}
To analyze the phonon properties and electronic density of states of the non-SC and SC samples, we have performed the heat capacity measurements for the x = 0 and 0.3 (highest T$_{c}$) samples. \figref{Fig10:CP} illustrates the C$_{P}$/T as a function of T$^{2}$ above 1.8 K. The superconducting transition, which is manifested by a jump in the heat capacity data is observed at 4 K, confirming the bulk superconductivity in Mo$_{0.7}$Re$_{0.3}$Te$_{2}$. The low-temperature normal state specific heat can be extracted using the equation $C_{P}/T$ = $\gamma$ + $\beta$$T^{2}$ ($\gamma$ = Sommerfeld coefficient and $\beta$ is the lattice contribution to the specific heat) which yields $\gamma$ = 4.7 and 9.0 mJ/mol K$^{2}$ for x = 0 and 0.3 respectively. Considering the free electron model, the density of states values at the Fermi level [$N(E_{F})$ =  $\frac{3\gamma}{\pi^{2}K_{B}^{2}}$] is calculated as 2.01 and 3.83 states eV$^{-1}$ f.u.$^{-1}$ for x = 0 and 0.3 respectively. The electron-phonon coupling constant, which gives the strength of the attraction between the electron and phonon can be calculated using the McMillan formula \cite{37} 
\begin{equation}
\lambda_{e-ph} = \frac{1.04+\mu^{*}ln(\theta_{D}/1.45T_{c})}{(1-0.62\mu^{*})ln(\theta_{D}/1.45T_{c})-1.04 }
\label{eqn3:ld}
\end{equation} 

By considering the Coulomb pseudopotential $\mu^{*}$ = 0.1 \cite{29}, $\lambda_{e-ph}$ are calculated as 0.31 and 0.64 for x = 0 and x = 0.3, respectively. These values indicate that the compounds are weak coupling superconductors. The increase in the electron concentration (by substituting the Re for Mo in MoTe$_{2}$) may facilitates the enhancement of the electron-phonon coupling and DOS at the Fermi energy. This, in fact, indicates that a significant reconstruction of the Fermi surface is occurring with Re substitution. The elementary density functional theory calculations \cite{38} also indicates the enhancement of DOS. However, further experiments are required to gain deeper insights into Fermi surface reconstruction. The electron doping by means of Re substitution in the MoTe$_{2}$ phase shows the emergence of superconductivity, whereas the hole doping (Nb-doped MoTe$_{2}$ \cite{31}) in MoTe$_{2}$ phase does not show superconductivity.

\section{Conclusion}

In conclusion, we have successfully prepared a series of Mo$_{1-x}$Re$_{x}$Te$_{2}$ samples. Similar to the pristine 1T$'$ MoTe$_{2}$, at room temperature all the samples crystallize in a CdI$_{2}$ type centrosymmetric monoclinic structure (P21/m space group) consisting of edge-sharing (Mo/Re)Te$_{6}$ octahedra. The magnetic, transport and specific heat measurements highlight a dramatic enhancement of the superconducting transition temperature with Re substitution. Combining specific heat, Hall measurements indicate that the Re substitution is doping electrons. It can be a key reason to facilitates the emergence of superconductivity by enhancing the electron-phonon coupling and DOS at the Fermi energy. The estimated upper critical field (H$_{C2}$(0) = 7.2 T) for the Re rich composition Mo$_{0.6}$Re$_{0.4}$Te$_{2}$ is almost equal to the BCS weak coupling Pauli limit (7.23 T). It may be the indication of unconventional superconductivity. However, further experiments (such as Muon spectroscopy) are required to verify this point. A T$_{c}$, as high as 4.1 K is observed for the Mo$_{0.7}$Re$_{0.3}$Te$_{2}$ composition. In our best of knowledge, this is the highest T$_{c}$ at ambient pressure, achieved yet in a 1T$'$ MoTe$_{2}$ related sample. Unlike the pressure-enhanced superconducting state in Weyl semimetal MoTe$_{2}$ (and/or WTe$_{2}$), our findings open a new way to further manipulate and enhance the superconducting state together with the topological states in 2D van der Waals materials. 

\section{Acknowledgments}

R.~P.~S.\ acknowledges Science and Engineering Research Board (SERB), Government of India for the Young Scientist Grant No. YSS/2015/001799 and a Ramanujan Fellowship through Grant No. SR/S2/RJN-83/2012. S.~M.\ acknowledges SERB, Government of India for the NPDF Fellowship (PDF/2016/000348). N.~G.\ acknowledges SERB Grant No. ECR/2016/001004 and the use of HPC facility of IISER Bhopal. Financial support from DST-FIST Project No. SR/FST/PSI-195/2014(C) and DST-Nano Mission is also thankfully acknowledged 

\section{Contributions}

M.~M.\ and S.~M.\ contributed this work equally.

\end{document}